\documentclass[twocolumn,showpacs,preprintnumbers,amsmath,amssymb,floatfix,aps]{revtex4-1}

\usepackage{graphicx}
\usepackage{amssymb}
\usepackage{epstopdf}
\usepackage{float}
\usepackage{color}
\usepackage{subfigure}
\usepackage{multirow}
\usepackage{relsize}
\usepackage{booktabs}

\begin{document}
	
\preprint{APS/123-QED}
	
\title{Short-range Photoassociation of LiRb}

\author{D. Blasing$^{1,*}$, I. C. Stevenson$^{2,}$} 
\thanks{These two authors contributed equally}
\affiliation{%
$^1$Department of Physics and Astronomy, $^2$School of Electrical and Computer Engineering, and $^3$Purdue Quantum Center, Purdue University, West Lafayette, IN 47907
}%
\email[Corresponding author: David Blasing, ]{dblasing@purdue.edu}

\author{J. P\'{e}rez-R\'{\i}os$^1$, D. S. Elliott$^{1,2,3}$, and Y. P. Chen$^{1,2,3}$} 
\affiliation{%
$^1$Department of Physics and Astronomy, $^2$School of Electrical and Computer Engineering, and $^3$Purdue Quantum Center, Purdue University, West Lafayette, IN 47907
}%

\date{\today}% It is always \today, today,
	
\begin{abstract}
We have observed short-range photoassociation of LiRb to the two lowest vibrational states of the $d\,^3\Pi$ potential. These $d\,^3\Pi$ molecules then spontaneously decay to vibrational levels of the $a^3\,\Sigma^+$ state with generation rates of $\sim10^3$ molecules per second. This is the first observation of many of these $a\,^3\Sigma^+$ levels. We observe an alternation of the peak heights in the rotational photoassociation spectrum that suggests a $p$-wave shape resonance in the scattering state. Franck-Condon overlap calculations predict that photoassociation to higher vibrational levels of the $d\,^3\Pi$, in particular the sixth vibrational level, should populate the lowest vibrational level of the $a\,^3\Sigma^+$ state with a rate as high as $10^4$ molecules per second. These results encourage further work to explain our observed LiRb collisional physics using PECs. This work also motivates an experimental search for short-range photoassociation to other bound molecules, such as the $c\,^3\Sigma^+$ or $b\,^3\Pi$, as prospects for preparing ground-state molecules.
\end{abstract}

\maketitle

%\section{Introduction}

%Ultracold polar molecules have been of interest for some time~\cite{DeMille2002,Ni2010,Krems2009}.  The permanent electric dipole moment of these molecules in their rovibrational ground state allows a range of unique physical interactions~\cite{Trefzger2009,Rabl2007} that are inaccessible to ultracold atomic systems.  The most recent addition to the bi-alkali family of ultracold molecules is $^7$Li-$^{85}$Rb~\cite{Dutta2013,Dutta2014,Altaf2015,Lorenz2014}.  This molecule is characterized by a large electric dipole moment~\cite{Aymar2005} (several times larger than that of KRb and RbCs, for example), and a high rate of photoassociation (PA).

Heteronuclear bi-alkali molecules in the $X\,^1\Sigma^+$ or $a\,^3\Sigma^+$ electronic potentials are interesting both experimentally and theoretically for a number of reasons. Experimentally, they have long lifetimes and large, permanent electric dipole moments. Significant effort has been dedicated to the study of dipolar molecules \cite{Baranov2008,Krems2009,Lahaye2009}, in part because permanent electric dipole moments give rise to interesting long-range and anisotropic interactions \cite{Ni2010}. For example, dipolar bosons may exhibit a pair supersolid phase \cite{Trefzger2009}, and enhance \cite{Gorshkov2011} or destabilize \cite{Zhang2015} superfluidity. They have been proposed as qubits for quantum computation \cite{Yelin2006,DeMille2002}, which, under certain conditions, could exhibit high fidelity \cite{Rabl2007}. Trapped ensembles of heteronuclear bi-alkali molecules could also exhibit novel few-body \cite{Buchler2007} and many-body interactions \cite{PerezRios2010,Wang2006,Micheli2006}. Such molecules could even be used to probe for variation of fundamental constants \cite{Chin2009}. 

The first step in any experimental realization is, of course, the creation of the ultracold heteronuclear bi-alkali molecules themselves. Two preparation methods stand out in particular: magnetoassociation followed by Stimulated Raman Adiabatic Passage (STIRAP) \cite{Ni2008}, and photoassociation (PA) followed by spontaneous emission \cite{Deiglmayr2008,Zabawa2011}. The PA method is experimentally simpler (as it only involves one laser), but it relies on finding an excited state that decays preferentially to the desired final state. To extend the study of the rich physics offered by ultracold heteronuclear bi-alkali molecules, various preparation methods must be evaluated in a variety of systems. In this work using ultracold LiRb, we have evaluated one pathway to create $a\,^3\Sigma^+$ molecules, namely PA of atoms to the $d\,^3\Pi$ molecular state. 

Photoassociation is the process where unbound atoms, colliding in the presence of light, can absorb a photon and bind into an electronically excited state molecule. In order to have a significant probability for this process to occur, the matrix element between the scattering atoms and the excited molecule through the dipole operator must be large. This matrix element is typically large, and thus PA often occurs, for excited state molecules with large internuclear separation. It can also sometimes be large at short-range, and short-range PA has been seen in LiCs \cite{Deiglmayr2008}, NaCs \cite{Zabawa2011}, and RbCs \cite{Gabbanini2011,Ji2012,Bouloufa-Maafa2012,Bruzewicz2014}. New short-range excited state molecules are interesting and useful to study because they can decay to deeply bound vibrational levels in the $X\,^1\Sigma^+$ or $a\,^3\Sigma^+$ electronic potentials \cite{Deiglmayr2008,Zabawa2011,Banerjee2012,Bruzewicz2014,Stevenson2016}. In the present work, we report on short-range PA to the lowest vibrational levels in the $d \,^3\Pi$ electronic potential of LiRb. The $d \,^3\Pi$ molecules subsequently spontaneously decay to the $a\,^3\Sigma^+$ molecules (this is similar to that observed in RbCs \cite{Bruzewicz2014}). We generate molecules bound in the $a\,^3\Sigma^+$ potential with a rate of $\sim10^3$ molecules per second in the seventh vibrational state. We also predict a possible extension of our work that may generate $\sim2\times10^4$ molecules per second in the lowest vibrational state of the $a\,^3\Sigma^+$ state. A generation rate on the order of $10^4$ would rank among the highest of rates for the heteronuclear bi-alkali molecules in the lowest vibrational level.    

\begin{figure}[t]
	\includegraphics[width=8.6cm,trim= 0cm 0cm 27cm 0cm,clip=true]{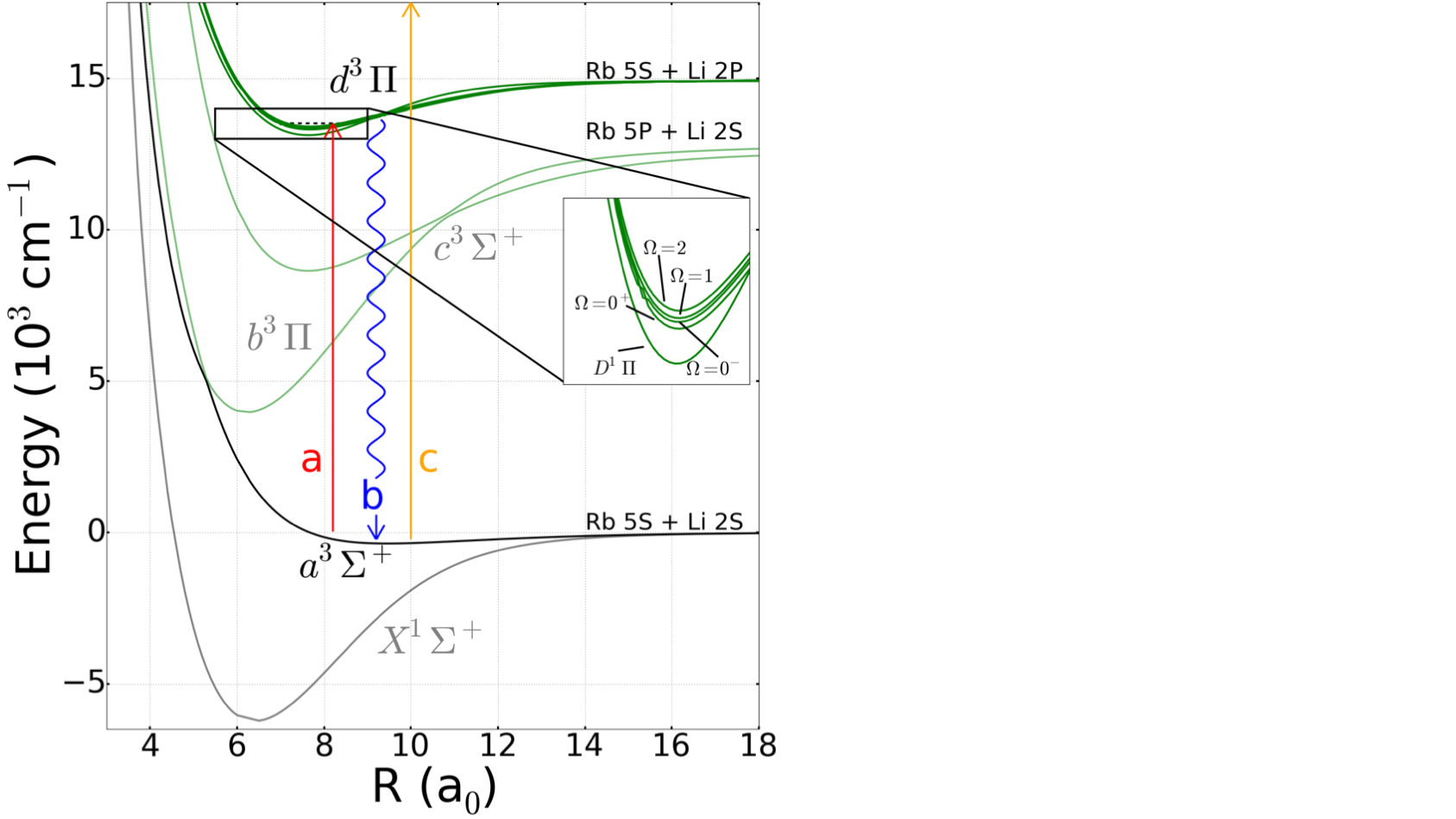}\\
	\caption{(Color on-line) Energy level diagram for the LiRb molecule showing the relevant PECs from Ref.~\cite{Korek2009}.  Vertical lines show transitions, including {\bf a} the photoassociation, with frequency $\nu_{a}$; {\bf b} spontaneous decay of excited state molecules leading to $a\,^3 \Sigma ^+$ molecules; {\bf c} first step of REMPI ionization of $a\,^3\Sigma^+$ molecules, with frequency $\nu_{c}$, through $f\,^3\Pi$ (PEC not shown for clarity). The horizontal black dashed line represents our PA states. Inset: The spin-orbit components of the $d\,^3\Pi$ state and also the nearby $D\,^1\Pi$ state.}
	\label{fig:PEC}
\end{figure}

The details of our experimental apparatus are contained in Ref. \cite{Dutta2014num2}, so here we provide only an overview.  We work out of a dual-species LiRb MOT with temperature $\lesssim\,1\,$ mK and diameter $\sim$ 1 mm ~\cite{Altaf2015}. We trap $\sim 5\times10^7$ Li atoms, and $\sim 1 \times 10^8$ Rb atoms, both primarily in their $F=2$ levels.  The Rb MOT is a spatial dark SPOT MOT~\cite{Ketterle1993}.  Photoassociation of Li and Rb atoms into $d\,^3\Pi$ molecules is driven by an $\sim$ 100 mW cw Ti:Sapphire laser. To explore the final vibrational state distribution of the molecules following their spontaneous emission, we ionize them using resonantly-enhanced-multi-photon-ionization (REMPI). The REMPI process is driven by a Nd:YAG pumped, pulsed dye laser. Its frequency, $\nu_c$, is tunable from 17150 to 18150 cm$^{-1}$ when using the R590 dye, and in a 4 mm diameter beam it delivers $\sim$1.5 mJ/pulse to the MOTs at a 10 Hz repetition rate.  When $\nu_c$ is resonant between an initial $a\,^3\Sigma^+$ state and an intermediate $f\,^3\Pi$ state, absorption of an additional photon at frequency $\nu_c$ can ionize the molecule. Then the LiRb$^+$ molecular ion is accelerated with a dc electric field into a microchannel plate detector for time-of-flight based counting. In this paper, $v$ and $J$ denote the vibrational and rotational levels of the $d\,^3\Pi_\Omega$ PA resonances, $v^{\prime}$ the vibrational states of the $f\,^3\Pi$ states used for REMPI, and $v^{\prime \prime}$ and $J^{\prime \prime}$ the vibrational and rotational levels of the $a\,^3 \Sigma ^+$ states which result from spontaneous decay. $\Omega$ is the projection along the internuclear axis of the total electronic angular momentum (i.e. orbital plus spin), and vibrational levels are counted up from lowest bound state. Fig. \ref{fig:PEC} shows the relevant frequencies, states, and potential energy curves (PECs).

\begin{figure}[t]
	\includegraphics[width=8.6cm,trim= 0cm 6.5cm 0cm 0cm,clip=true]{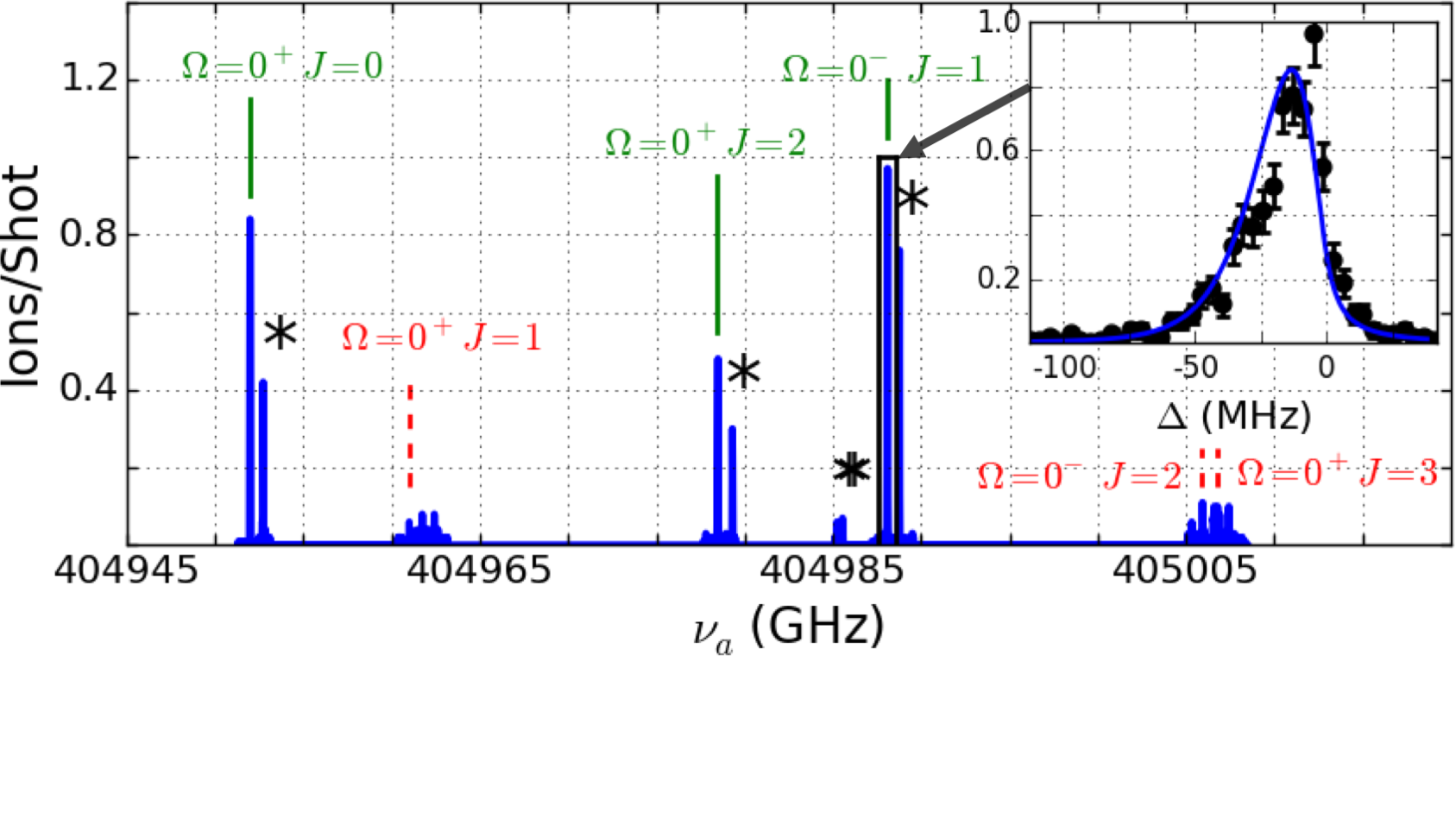}\\
	\caption{(Color on-line) PA spectrum of the $d\,^3\Pi \,\,v=0$ state.  We observe PA transitions from $J$ = 0 to 3, and have labeled PA to states with $\pm$ inversion parity as respectively green solid and red dashed lines. This parity alternates based on the $J$ quantum number in the $\Omega=0^\pm$ states. These assignments are consistent with those determined by depletion spectroscopy to the same states \cite{Stevenson2016}. We observe that free-to-bound PA transitions to molecular states with $+$ inversion parity are significantly stronger than their $-$ parity counterparts, sometimes by almost an order of magnitude. Inset: a high resolution PA scan to the $\Omega=0^-, J=1$ state with a fit based on Eq. (\ref{eq:fit}), $\Delta$ is the detuning from the fitted peak center. ($\ast$ denotes hyperfine echo from population in Li $F$ = 1 or Rb $F$= 3 atoms.)}
	\label{fig:paExample}
\end{figure}

We show a subset of the PA spectrum in Fig. \ref{fig:paExample}, and, in Table \ref{tab:pa_assignments}, we list the frequencies for all the PA resonances to $d\,^3\Pi$ molecules that we have observed. These include the first and second vibrational levels and their four spin-orbit split states ($\Omega=0^+$, $0^-$, $1$, and $2$). The most interesting feature is the strong enhancement of the PA rate to positive inversion parity $d\,^3\Pi$ states compared to those of negative inversion parity. We do not know of any previous bi-alkali experiment that observed such an alternation of PA rates. (For example, in LiCs ~\cite{Deiglmayr2008}, Rb$_2$ ~\cite{Bellos2011} or RbCs~\cite{Bruzewicz2014,Bouloufa-Maafa2012,Ji2012}, PA to adjacent rotational states had similar strengths within a factor of 2.) Within an $\Omega$ progression, the green and red labels that alternate for increasing $J$ refer to such parity of the PA state upon inversion through the origin. (The $\pm$ label for the $\Omega=0^\pm$ states refers to a different parity, that of reflection through a plane containing both nuclei.) For $\Omega=0^+$, the inversion parity is $(-1)^{\mathrm{J}}$, while for $\Omega=0^-$, it is $(-1)^{\mathrm{J+1}}$. For dipole transitions, the parity of the initial and final states must be opposite. Therefore, a possible explanation of our higher PA rates to states with positive inversion parity is that, at the temperature of our MOTs, a partial wave with negative inversion parity makes the largest contribution to the collision cross-section between Li and Rb. The parity of the $\ell^{\text{th}}$ partial wave is $(-1)^{\ell}$, where $\ell$ is the angular momentum quantum number of the collision; the lowest partial wave with negative parity is the $\ell=1$ ($p$-wave).

\begin{table}[h]
	\centering
	\begin{tabular}{c c c c c c c c c c }
  \toprule[1pt]
		 $v$ &$\Omega$ & J=$0$ & J=$1$ & J=$2$ &J=$3$& $B_{v}$\\
		\hline
		0& 0$^+$ & 404952.0 & 404961.0 & 404978.5 & 405006.8 & 4.5 \\
		%\hline
		0&0$^-$ &  & 404988.1 & 405005.9 &  & 4.5 \\
		%\hline
		0&1 & ------ & 406062.7 & 406080.6 & 406108.4 & 4.5 \\
		%\hline
		0&2 & ------ & ------ &  407067 &  & \\
		%\hline
		1&0$^+$ & 407918.6 & \footnote{We observed a peak in the PA spectrum at 407928.2 GHz, which is 800 MHz higher than the predicted value of 407927.4 GHz using the rotational constant of 4.4 GHz (as determined by the other peaks in the series). The identity of this peak is unclear since 800 MHz is also our Li hyperfine spacing.} & 407944.9 &  & 4.4 \\
		%\hline
		1&0$^-$ &  & 407952.0 &  &  & \\
		%\hline
		1&1 & ------ & 409037.4 & 409054.9 &  & 4.4\\
		\bottomrule[1pt]
	\end{tabular}
	\caption{The frequencies for the observed $d\,^3\Pi$ PA resonances in GHz. Uncertainties are $\pm2$ in the last digit recorded. Blank entries denote allowed transitions that did not appear in our spectra; solid horizontal lines denote forbidden transitions. The $v=0$ splitting of the spin-orbit levels are $27.7,1074.2,$ and $986.4,$ (all $\pm .3$) GHz for $\Omega=0^-/0^+,1/0^-,$ and $2/1$, respectively, which differ significantly from their predicted values \cite{Korek2009}. The $J$-dependent inversion symmetry for $\Omega\neq0$ states was not resolved since the $\Lambda$ doubling for the low lying rotational states accessed is small. The additional uncertainty for the $d\,^3\Pi_2,v=0,J=2$ line position is due to its significantly lower PA strength and its more complicated structure. The spin-orbit splittings and rotational constants are in agreement with our recent measurements using depletion spectroscopy \cite{Stevenson2016}.}
	\label{tab:pa_assignments}
\end{table}

\begin{figure}[t]
	\includegraphics[width=8.6cm,trim=1cm 0cm 7.5cm 0cm,clip=true]{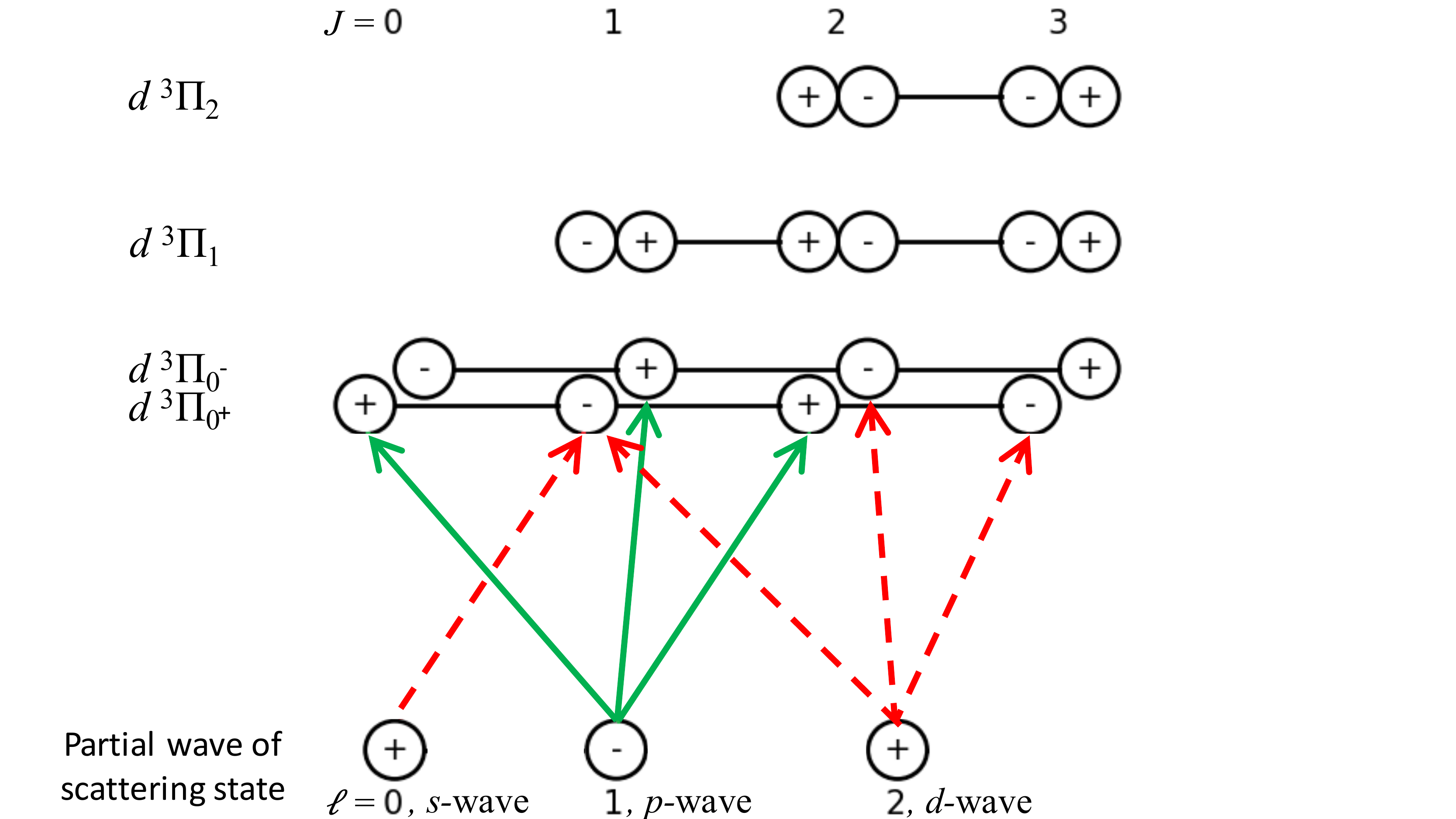}\\
	\caption{(Color on-line) Inversion parities for the partial waves of the scattering state and final $d\,^3\Pi_\Omega$ states, adapted and modified from Ref. \cite{Herzberg1989}. The solid green and dashed red arrows correspond respectively to the alternating strong and weak dipole-allowed PA transitions to the $d\,^3\Pi_{0}$ shown in Fig. \ref{fig:paExample}. The parity of every state or wave is shown as a plus or minus sign inside the circles. The transitions to $\Omega=$1 or 2 states and the small energy splitting of the $J$ states are not shown for clarity.  Note that we can resolve the energy splitting between different parities for the $\Omega = 0^\pm$ states, but not for $\Omega=$1 or 2 states.}
	\label{fig:parity}
\end{figure}
%\begin{figure}[h]
%	% Requires \usepackage{graphicx}
%	\includegraphics[scale=0.4]{xs_figure.eps}\\
%	\caption{(Color on-line) Elastic cross section as a function of the collision energy for $^7$Li-$^85$Rb colliding throughout the $a^3\Sigma^{+}$ PEC. The inset shows the partial waves associated with different partial waves from $s$-wave up to $g$-wave contributing to the elastic cross section.}
%	\label{fig:crosssection}
%\end{figure}

%Further, the total angular momentum of a molecule resulting from PA is $\vec{J} = \vec{\ell} + \vec{j_a} + \vec{j_b}$, where $\vec{\ell}$ is quantum number for the scattering partial wave, and $\vec{j_a}$ and $ \vec{j_b}$ are the total angular momentum of the atomic asymptotes of the excited molecular state\cite{Jones1999}.  The states $d\, ^3\Pi_{0^-}$ and $d\,^3\Pi_{1}$ are asymptotic to Rb 5S$_{1/2}$ + Li 2P$_{1/2}$, while the states $d\,^3\Pi_{0^+}$ and $d\,^3\Pi_{2}$ are asymptotic to Rb 5S$_{1/2}$ + Li 2P$_{3/2}$~\cite{Korek2009}. Thus, for the state $d\, ^3\Pi_{0^-}$, $j_a = j_b = 1/2$. For this state, adding the angular momenta, and accounting for the parity of the initial scattering state along the $a\,^3\Sigma^+$ potential and the final $d\,^3\Pi$ bound states yields the following. The $s$-wave channel contributes to J = 0 or 1 with negative parity, while $p$-wave scattering will contribute to $J=0,1$ or $2$ with positive parity. $d$-wave could produce a negative parity $J=1,2,$ or $3$ and $f$-wave a positive parity J=$1,2,3,$ and $4$. 

Several aspects of our data support the speculation that our scattering state has significant $p$-wave nature at the temperature of our MOTs. First, in the present work, we observe a PA rate that alternates for J states in the $\Omega$ = 0$^\pm$, but not for $J=$ 1 and 2 of the $\Omega$ = 1. For $\Omega$ = 0$^\pm$, different $J$ states alternate in parity but for  $\Omega$ = 1, both parities are available at each $J$ state \cite{Herzberg1989}. Since positive parity excited states have significantly larger PA rates, we conclude that the scattering state has predominantly negative parity. Also, since PA to $J=3$ of either $\Omega=0^-$ or $0^+$ was either weak or not observed, the negative wave of the scattering state is likely to be the $p$-wave (and not the $f$-wave). Second, in a different set of measurements, we were able to PA to even-parity states of other electronic states \cite{Stevenson2016}. Both observations are consistent with the $p$-wave being significant for our scattering state, and in Fig. \ref{fig:parity} we show our interpretation for the different partial waves responsible for the PA spectrum in Fig. \ref{fig:paExample}.

To further understand if $p$-wave scattering may dominate the collisional physics, we first determined the temperature, $T,$ of the collisions between Li and Rb in our MOTs. This was done by fitting the experimental line shape of the PA resonances with a convolution of a Boltzmann with a Lorentzian \cite{Jones1999}:

\begin{equation}
\label{eq:fit}
W(f,f_{0})\propto \sum_l \int_{0}^{\infty}e^{-\frac{h\nu}{k_{\text{B}}T}} \nu^{\ell+\frac{1}{2}} L_{\Gamma}(f,f_{0}-\nu)d\nu,
\end{equation}

\noindent
where $L_{\Gamma}(f,f_{0})$ stands for a Lorentzian function with central frequency $f_{0}$ and natural line width $\Gamma$, and $k_{\text{B}}$ and $h$ denote the Boltzmann and Planck constants respectively. We fit our PA resonances with Eq. (\ref{eq:fit}) using four free parameters: the overall amplitude, the natural linewidth $\Gamma$, the resonant frequency $f_{0}$, and the collision temperature $T$. For example, the inset of Fig. \ref{fig:paExample} shows the PA spectrum for the $J$ = 1 level of the $\Omega$ = 0$^-$ electronic state and its fit. The fit assuming $p$-wave scattering yielded $T=360\pm 90\, \mu$K and $\Gamma= 8.6\pm 1.5$ MHz. (The data can also be fit with $\ell=0$, $s$-wave scattering, yielding similar temperatures and comparable fit quality measured by $R$-squared values. So $\ell$ for the scattering state cannot be determined from the fits.)

We have used the currently available PECs for LiRb \cite{Korek2009,Ivanova2011,Maier2015} to calculate the contributions to the elastic scattering cross-section of the few lowest partial waves at various temperatures including $360\,\,\mu K$. The calculations predicted that the $s$-wave, not the $p$-wave, is dominant at 360 $\mu$K. However, such calculations depend very sensitively on the PEC. Therefore, resolving the apparent disagreement between our interpretation of the experimental results and the theoretical PEC-based calculations of the scattering cross-section is a worthwhile, open question. (The PECs were sufficient to derive the binding energies of most of the vibrational levels reasonably well and a scattering length consistent with those previously predicted \cite{Marzok2009,PerezRios2015}.)

%One mechanism that would enable $p$-wave scattering to dominate below the centrifugal barrier at 1.9 mK~\cite{Altaf2015} is a $p$-wave shape resonance in the Li-Rb scattering channel. Indeed, similar shape resonances have been observed in LiCs ($d$-wave)~\cite{Deiglmayr2008} and NaCs ($f$-wave)~\cite{Zabawa2011}. In order to clarify this point further we have conducted single-channel elastic scattering calculations for Li-Rb system in the $a^3\Sigma^{+}$ PEC (potential energy curve) employing the analytical PEC of Ivanova et al. \cite{Ivanova2011} and the results are shown in Fig. \ref{fig:crosssection}. In this figure it is noticed that around 1.5 mK the $p$-wave scattering becomes the dominant contribution to the cross section, but there isn't any trace or indication of a $p$-wave shape resonance. However we found a $g$-wave shape resonance around 100 mK. These results seems to indicate that $-$ parity PA states should have a strongest signal, which is the opposite to what we observe. This suggest that the $a^3\Sigma^{+}$ employed in the calculations may require some modifications in order to capture this fine detail of the Li-Rb interaction, which can be crucial for future studies in Feshbach molecular formation in the system at hand.

%\section{Spontaneous decay to the $\bf{a^3\Sigma^+\,v^{\prime\prime}}$ manifold}
%\label{sec:spontdecay}

\begin{figure}[t]
	\includegraphics[width=8.6cm,trim=6cm 0cm 5.5cm 2cm,clip=true]{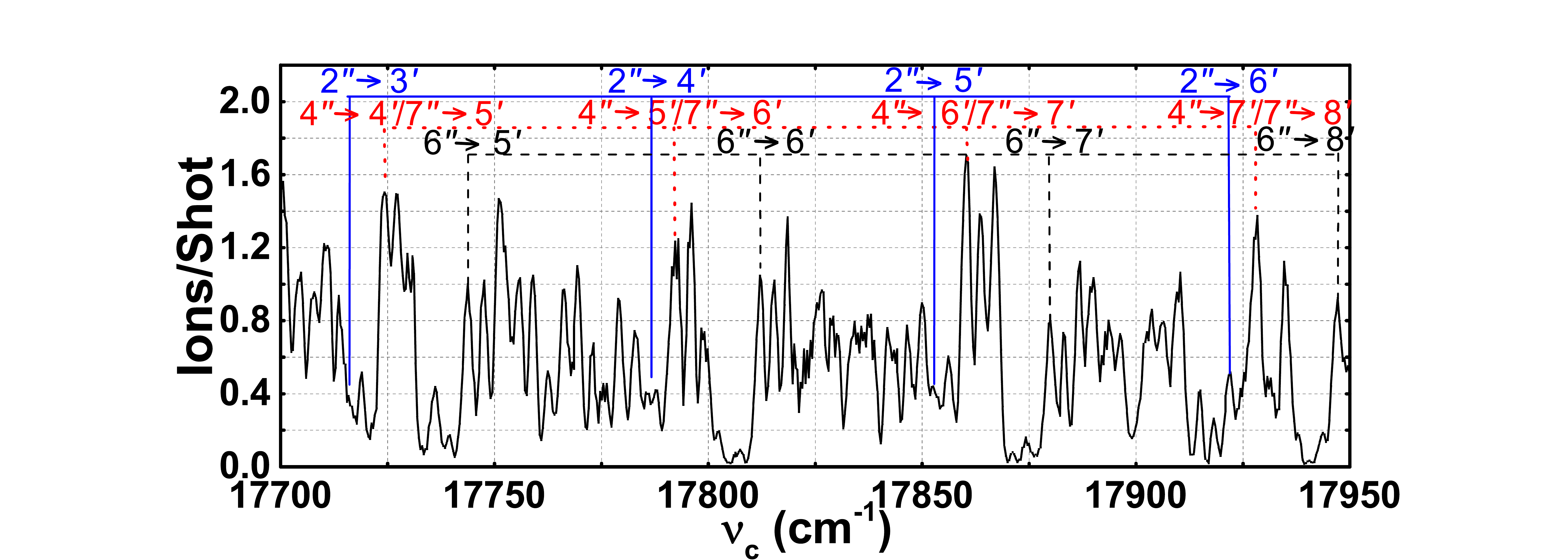}\\
	\caption{(Color on-line) Sample REMPI spectrum. The blue solid, red dotted, and black dashed lines respectively are our assignments for progressions from $a\,^3\Sigma^+ \,\,v^{\prime\prime}=2,$ some combination of $v^{\prime\prime}=$ 4 or 7, and $v^{\prime\prime}=6$ to vibrational levels $v^\prime = 3-8$ of the $f\,^3\Pi_0$ state. Each data point is the average of 100 REMPI pulses, and is smoothed by averaging over nearest neighbors. (Although not labeled in the figure, we have assignments for more than 90\% of observed REMPI resonances.)}
	\label{fig:REMPIexample}
\end{figure}

After exploring the properties of the PA to the $d\,^3\Pi$, we investigated the spontaneous decay of the $d\,^3\Pi_{0^-},v=0,J=1$ state with REMPI spectroscopy. For this spectroscopy we locked the Ti:Sapphire laser to the PA resonance of the $d\,^3\Pi_{0^-},v=0,J=1$. The motivation for this was to explore deeply bound molecules of the $a\,^3\Sigma^+$ potential, perhaps even its lowest vibrational level.  We show a sample REMPI spectrum, with transitions from $a\,^3\Sigma^+\,v^{\prime\prime}=2,4,6,$ and $7$ in Fig. \ref{fig:REMPIexample}. In previous studies, we have used PA (to other bound states) followed by REMPI spectroscopy to detect $a\,^3\Sigma^+\,v^{\prime\prime}=7\rightarrow13$ levels \cite{Altaf2015}. In the current study, and from the whole REMPI spectrum, the progressions for $v^{\prime\prime}=2,4,6,7,8,9,$ and $10$ are reasonably clear. However, ionizing $a\,^3\Sigma^+$ states $v^{\prime\prime}=1$ and $4$ through the $f\,^3\Pi$ states results in congested REMPI peaks and our assignments are less certain.

From the REMPI spectrum and our previous spectroscopy of the $f\,^3\Pi_0$ states \cite{Altaf2015}, we extracted binding energies for some new, lower vibrational levels of the $a\,^3\Sigma^+$ states as listed in Table \ref{tab:bindingenergies}. Further experiments, particularly depletion spectroscopy, would refine our line assignments and binding energies. 

%Although not assigned, our data is partially, though not fully, consistent with decay to $v^{\prime\prime}=$ 0 and 3 \footnote{It is our udgement that the data does not clearly demonstrate a $v^{\prime\prime}=0$ series. There was an incomplete series that, if more regular, we would have assigned to $v^{\prime\prime}=0\rightarrow f\,^3\Pi_{\Omega^{\prime}}\,v^\prime$ with a binding energy around 273 cm$^{-1}$. The REMPI peaks from the $v^{\prime\prime}=3\rightarrow f\,^3\Pi\,v^\prime$ series were never unambiguously seen (strong peaks from other $a\,^3\Sigma^+\, v^{\prime\prime}\rightarrow f\,^3\Pi_{\Omega^{\prime}}\,v^\prime$ transitions always obfuscated the small peaks that could have been assigned to $v^{\prime\prime}=3$). There was inconclusive evidence for a state with binding energy between 154 cm$^{-1}$ and 162 cm$^{-1}$.}. 

\begin{table}[h]
	\centering
	\begin{tabular}{c c c c c c c c c}
		  \toprule[1pt]
		\multirow{2}{*}{$a\,^3\Sigma^+\,\, v^{\prime\prime}$} &  \multicolumn{2}{c}{E$_\mathrm{B}$ (cm$^{-1}$)\,\,}& \multicolumn{2}{c}{E$_\mathrm{B}(v^{\prime\prime})$-E$_\mathrm{B} (v^{\prime\prime}+1)$ (cm$^{-1}$)} \\ 
				 &  \multicolumn{2}{c}{Exp.\,\,\,\,\,\,\,Theo.\,\,\,\,}& \multicolumn{2}{c}{\,\,Exp.\hspace{1.1cm}Theo.}  \\
		\hline
		$0$ & & 257.7  &  & 37.7  \\
		$1$ & 222 & 220.0  & \hspace{.55cm}34.9\hspace{.55cm} & 34.6  \\
		$2$ & 186.7 & 185.4  &  & 31.7  \\
		$3$ &  & 153.7  &   &  28.7  \\
		$4$ & 126 & 125 & 26.1 & 25.7  \\
		$5$ &  99.8 & 99.3 & 23 & 22.7  \\
		$6$ & 76.8 & 76.5  & 19.6 & 19.8  \\
		$7$ & 57.2 & 56.8 & 16.6 & 16.7  \\
		$8$ & 40.6 & 40.1 & 13.9 & 13.6  \\
		$9$ & 26.7 & 26.5 & 10.2 & 10.4  \\
		$10$ & 16.5 & 16.1 &  & 7.4  \\
		$13$ & 1.08~\footnote{This value was extracted from depletion spectroscopy using a different PA resonance \cite{Stevenson2016} in combination with the PA data presented here and has an uncertainty of 0.02 cm$^{-1}$; the higher precision of this binding energy allowed us to extract the other binding energies at the $\sim.1\rightarrow1$ cm$^{-1}$ level.} & 1.1 & & 1.0 &  \\
		\bottomrule[1pt]
	\end{tabular}
	\caption{Binding energies (E$_\mathrm{B}$) of $a\,^3\Sigma^+\,\, v^{\prime\prime}$ states extracted from our REMPI spectrum. The binding energies for $v^{\prime\prime}=2,5,6,7,8,9,$ and $10$ have uncertainties of $\pm$ .2 cm$^{-1}$. The binding energies for $ v^{\prime\prime}=1$ and $4$ have uncertainties of $\pm$ 1 cm$^{-1}$ since their REMPI lines occurred in regions with significant line congestion. The blank entries correspond to states and binding energies which we were not able to determine from the REMPI spectrum. The theoretical predictions use the potential energy curves from Refs. \cite{Korek2009,Ivanova2011} with the LEVEL 8.0 code \cite{Roy2012}.}
	\label{tab:bindingenergies}
\end{table}

For many future uses of polar molecules, the rate, $R$, of generating molecules is important. To estimate $R,$ we follow a similar procedure to that used in Ref. \cite{Stevenson2016_v0}. For the strongest and weakest REMPI lines, assigned to $a\,^3\Sigma^+\,v^{\prime\prime}=7$ and $v^{\prime\prime}=2$, we find rates of $\approx 2.5\times10\,^3$ molecules and $\approx 4\times10^2$ per second, respectively. The generation rates of the other vibrational levels of the $a\,^3\Sigma^+$ fall between those values.  We are currently building a 1064 nm optical dipole trap to confine both Li and Rb. This will increase the densities of both species, and reduce both their temperatures, and thus lead to a higher PA rate.

\begin{figure}[!t]
	\includegraphics[width=8.6cm,trim=1cm 1cm 3cm 2cm,clip=true]{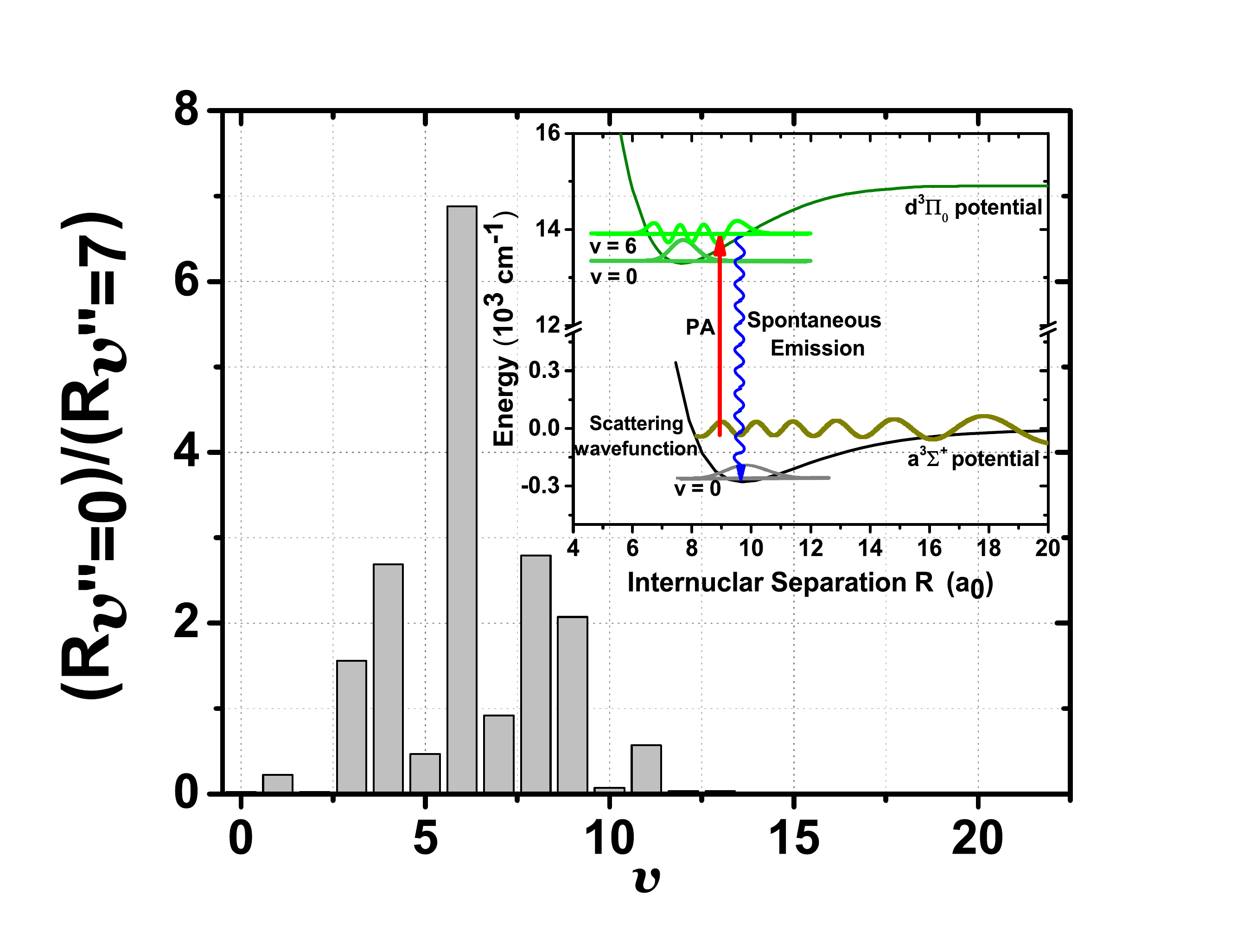}\\
	\caption{(Color on-line). Predicted generation rates, $R$, of the $a\,^3\Sigma^+,v^{\prime\prime}=0$ state using PA to different vibrational levels of the $d\,^3\Pi$ states. The result is normalized to the predicted rate for $a\,^3\Sigma^+,v^{\prime\prime}=7$ from PA to the $d\,^3\Pi_{0^-}\,v=0$. The $v^{\prime\prime}=0$ yielding a low normalized rate is consistent with our inability to see it in our REMPI scans that used PA to$d\,^3\Pi_{0^-}\,v=0$. Inset: sketch of PECs, bound wavefunctions of interest, and the scattering wavefunction at $500\,\,\mu K$. Note the high generation rate of $v^{\prime\prime}$=0 predicted from PA using the $d\,^3\Pi\,v$=6.}
	\label{fig:ratea3sigmav0}
\end{figure}

Our REMPI spectrum did not conclusively reveal the $v^{\prime\prime}=0$ state of the $a\,^3\Sigma^+$ potential. Unfortunately, other stronger REMPI lines obscured any weak lines that may have originated from $v^{\prime\prime}=0$ molecules. A possible way to create significant numbers of $v^{\prime\prime}=0$ molecules is PA to the $v=6$ of the $d\,^3\Pi$ state. (Our Ti:Sapphire laser cannot produce frequencies for PA to levels higher than to $v=1$.) Using the Numerov method to calculate the free-to-bound overlap for the PA step up, and LEVEL 8.0 for the Frank-Condon factor for the spontaneous emission step down, we predicted $R$ for $a\,^3\Sigma^+\,\,v^{\prime\prime}=0$ molecules while using PA to various vibrational levels of the $d\,^3\Pi$. We plot in Fig. \ref{fig:ratea3sigmav0} the predicted $R_{\text{v}^{\prime\prime}=0}$ for PA to various vibrational states of the $d\,^3\Pi$, normalized to the predicted $R_{v^{\prime\prime}=7}$ for PA to $d\,^3\Pi\, v=0$ (we use $v^{\prime\prime}=7$ since it is the assignment for our largest REMPI line). We find that PA to $v=6$ of the $d\,^3\Pi$ may produce $a\,^3\Sigma^+\,\,v^{\prime\prime}=0$ molecules at $\approx7\times$ the rate of our measured generation of $v^{\prime\prime}=7$. In the past, we have found Frank-Condon factors to have semi-quantitative predictive value. Therefore, spontaneous decay from PA to $v=6$ of the $d\,^3\Pi$ may result in $R_{v^{\prime\prime}=0}$ as high as $\sim2\times10^4$ molecules per second, and also fix the $a\,^3\Sigma^+$ well depth. 

%\section{Conclusions}
%\label{sec:conclusions}

Our experiments open a number of avenues for further work. The relative strengths of our PA lines seemingly indicate that $p$-wave scattering significantly affects the collisional physics of LiRb at $\approx 360\mu K$, which we can not derive from the currently available PECs. Precise measurements of the binding energies of the vibrational levels of the $a\,^3\Sigma^+$ states, particularly its higher vibrational levels, may help refine the $a\,^3\Sigma^+$ PEC. This work also motivates a search for short-range PA to the $c\,^3\Sigma^+$ state and the $b\,^3\Pi$ state, perhaps finding an excited state that decays preferentially to a small number of vibrational levels in the $a\,^3\Sigma^+$ potential; the $v=6$ state of the $d\,^3\Pi$ may provide such preferential decay and form $a\,^3\Sigma^+, v^{\prime\prime}=0$ molecules at significant rates. As such, our experiment lays solid groundwork towards efficient preparation of triplet molecular samples and the further experiments that need them.

We acknowledge helpful discussions with and the work of: Sourav Dutta, Adeel Altaf, and John Lorenz. The experimental work was funded with the Purdue AMO incentive grant and J. P.-R. acknowledges support from NSF Grant No. PHY-130690.
 
\bibliography{shortRangePA_Bib}{}
\bibliographystyle{apsrev4-1}

\end{document}